\documentclass[a4paper,11pt]{article}
\pdfoutput=1 

\usepackage{jcappub,aas_macros} 

\usepackage[T1]{fontenc} 

\newcommand{\M}[0]{\mathrm{Mpc}/h}

\newcommand{\vk}[0]{\mathbf{k}}
\newcommand{\fnl}{f_{\rm NL}}
\newcommand{\fnlloc}{f_{\rm NL}}

\newcommand{\kv}{\mathbf{k}}
\newcommand{\xv}{\mathbf{x}}
\newcommand{\qv}{\mathbf{q}}

\newcommand{\bphi}{b_{\phi}}
\newcommand{\bphifnl}{b_{\phi}f_{\rm NL}}
\newcommand{\bdphi}{b_{\delta\phi}}

\usepackage{hyperref}
\usepackage{fontawesome} 

\title{\boldmath Local Primordial Non-Gaussian Bias at the Field Level}

\author[a,b,d,1,2]{James M. Sullivan,\note{Corresponding author.} \note{Brinson Prize Fellow}}
\author[c]{Shi-Fan Chen}

\affiliation[a]{Department of Astronomy, University of California, Berkeley, CA 94720, USA}
\affiliation[b]{Berkeley Center for Cosmological Physics, University of California, Berkeley, CA 94720, USA}
\affiliation[c]{Institute for Advanced Study, 1 Einstein Drive, Princeton, NJ 08540, USA}
\affiliation[d]{Center for Theoretical Physics, Massachusetts Institute of Technology, Cambridge, MA 02139, USA}

\emailAdd{jms3@mit.edu}
\emailAdd{sfschen@ias.edu}

\abstract{
Local primordial non-Gaussianity (LPNG) couples long-wavelength cosmological fluctuations to the short-wavelength behavior of galaxies.
This coupling is encoded in bias parameters including $b_{\phi}$ and $b_{\delta\phi}$ at linear and quadratic order in the large-scale biasing framework.
We perform the first field-level measurement of $b_{\phi}$ and $b_{\delta\phi}$ using Lagrangian bias and non-linear displacements from N-body simulations.
We compare our field level measurements with universality predictions 
and separate universe results, finding qualitative consistency, but disagreement in detail.
We also quantify the information on $\fnlloc$ available in the field given various assumptions on knowledge of $b_{\phi}$ at fixed initial conditions.
We find that it is not possible to precisely constrain $\fnlloc$ when marginalizing over $\bphi \fnlloc$ even at the field level, observing a 2-3X degradation in constraints between a linear and quadratic biasing model on perturbative field-level mocks, suggesting that a $\bphi$ prior is necessary to meaningfully constrain $\fnlloc$ at the field level even in this idealized scenario.
For simulated dark matter halos, the pure $\fnlloc$ constraints from both linear and quadratic field-level models appear biased when marginalizing over bias parameters including $b_{\phi}$ and $b_{\delta\phi}$ due largely to the $\fnlloc b_\phi$ degeneracy.
Our results are an important consistency test of the large-scale bias framework for LPNG and highlight the importance of physically motivated priors on LPNG bias parameters for future surveys. 
}

\begin{document}
\maketitle
\flushbottom

\section{Introduction}
\label{sec:intro}

Cosmology links astronomically-observed large-scale fluctuations with high-energy physics in the early universe.
Primordial non-Gaussianity (PNG) in the early universe imprints spatial correlations in the cosmic microwave background (CMB) fluctuations and large-scale structure (LSS), acting as a late-time window into inflationary physics \cite{2022arXiv220308128A,2019BAAS...51c.107M}.
A particularly interesting variant is local PNG (LPNG) \cite{komatsu_lpng,chen_png_review,ligouri_png_cmb_review,creminnelli_lpng_estimator,lpng_wands}: this type of PNG is forbidden for single-field inflation by a correlator consistency relation \cite{maldacena03,acquaviva_consistency,pajer_consistency}, and, as such, its detection would be a smoking gun for more exotic inflationary scenarios. The amplitude of local PNG, $\fnl$ (here, we always mean $\fnl = f_{\mathrm{NL}}^{\mathrm{loc}}$), has been constrained by Planck data to be close to zero through a CMB bispectrum analysis at a level of $\sigma(\fnlloc)=5.1$ \cite{PlanckPNG,fergusson_bispectrum_lpng}.

LSS observations, most immediately in the form of galaxy surveys, have the potential to exceed the statistical sensitivity of Planck on $\fnlloc$ \cite{spherex14,dePutterDore17,heinrich_spherex_bispectrum_24,green_forecast_sdb_png_24,SchmittfullSeljak,sullivan_23,ZeroBias,seljak_mt,2014arXiv1412.4671A,barreira_krause_23,hamaus_halo_forecast,giannantonio_forecast_lpng,carbone_lpng_bias_forecast,alonso_lpng_forecast,karagiannis_lpng_forecast,moradinezhad_dizgah_lim_png,floss_cov_forecast,fondi_lpng_mt_forecast,Dizgah21,brown_lpng_desi_forecast}. Unlike in the CMB, nonlinear dynamics and galaxy evolution preclude constraints on LPNG directly from the large-scale bispectrum of galaxies. In the language of perturbation theory, these nonlinearities manifest as free bias parameters which contribute to the bispectrum in a highly degenerate manner with the LPNG signal, thereby diluting the LPNG constraint (see e.g. refs.~\cite{Cabass22BOSS,DAmico22BOSS} for the state-of-the-art in bispectrum LPNG constraints). On the other hand, LPNG also produces a very particular scale-dependent bias in the galaxy field that cannot be generated from Gaussian initial conditions and can be detected even at the level of the 2-point function (power spectrum) on very large scales \cite{Dalal08,desjacques_fnl_review,DJS,assassi_renorm_png,baldauf_gr,luisa_nn_bias,Lazeyras22,matarrese_lpng_bias,sefusatti_lpng_bias,pillepich_lpng_bias,dsi_lpng_bias,afshordi_lpng_bias,scoccimarro_lpng_bias,grossi_halo_lpng,jeong_komatsu_lpng_bias,pat_lpng_bias,giannantonio_lpng_bias,verde_09_lpng_bias,schmidt_kamionkowski_lpng_bias,djs_11_lpng_bias,kendrick_gnl_lpng_bias,quijote_png_halo,bareeria_lpng_h1,pat_bphi_08,adame_lpng_bias_unit,hadzhiyska_abacuspng,lazeyras_AB_quadratic_halos_21}.
This signal has an amplitude proportional to the level of LPNG, $\fnl$, but is also proportional to an unknown coefficient $b_\phi$, which depends on the small-scale astrophysics of the galaxy sample under consideration.

The scale-dependent bias due to LPNG allows us to constrain the amplitude $b_{\phi} f_{NL}$ instead of $f_{NL}$. To date, the majority of galaxy clustering data analyses have made strong assumptions about the value of $b_{\phi}$ to obtain tight constraints on $f_{NL}$ \cite{MuellereBOSS, rezaie_png_analysis, CastoriaeBOSS,Slosar08,LeistedtSDSSphoto,2022JCAP...11..013B,Cabass22BOSS,DAmico22BOSS,caligari_eboss_23, mccarthy_xcorr_cmb}, even when such assumptions may not be warranted for the observed tracers in question \cite{2020JCAP...12..031B,Lazeyras22,2022JCAP...11..013B,2020JCAP...12..013B,reid10_assmblyhistory}. 
Given this state of affairs, it is natural to wonder if using observables beyond $n-$point functions can extract additional information about $\fnlloc$ beyond the large-scale $\bphifnl$ contribution in the presence of unknown galaxy nonlinearities\footnote{A parallel and more empirical strategy is to more precisely quantify the sizes of these nonlinearities through numerical simulations \cite{sullivan_21,ivanov_priors}, such that the effect of LPNG in the bispectrum can be carefully isolated.}.
Many non-$n$-point statistics have been recently used to attempt to extract $\fnlloc$ information without explicit priors on $\bphi$ 
and some significant gains in $\fnlloc$ information over e.g. the power spectrum, have been demonstrated in simulations \cite{peron_wst_fnlloc,kNN_png,quijote_png_alternative,quijote_png_hmf,yip_ph_png,biagetti_ph,jamieson_emulator}.
However, these statistics often involve strong mixing of scales, or quantities that are not directly measurable, making their interpretation and use somewhat challenging.

Alternatively, it is possible to extract all the in-principle accessible information in the observed galaxy overdensity directly at the field-level, while still retaining perturbative modeling elements including scale-separation \cite{marcel_halo_field,marcel_rsd_field,nguyen_field_level,elsner_eft_like,fabian_eft_like,fabian_eft_like_2,fabian_s8_eft,stadler_eft_field_rsd,andrija_eftlike,andrej_h1_field}. In this context, it is possible to explore questions of $\fnlloc$ information (and the related question of $\bphi$ priors) without regard to the information lost when compressing the field, e.g., into the galaxy power spectrum. Such exploration is the focus of this work. While it has recently been suggested based on analytic arguments that the available perturbative information at the field level is equivalent to that in $n$-point functions \cite{cabass_pt_field_level}, other numerical investigations seemingly show large improvements from incorporating field-level information \cite{andrej_h1_field, nguyen_field_level}, and it has also been suggested that the particular degeneracies of PNG with nonlinear dynamics can be broken in this way \cite{Baumann22}.

In this paper, we explore the feasibility of constraining $\fnlloc$ using a field level forward model. For the first time, we use a bias model including extra operators generated by LPNG at the field level. 
We briefly review these operators, describe our field level bias model, and the N-body simulations used to compute galaxy trajectories and halo catalogs in Section~\ref{sec:model}.
We present bias parameter measurements for simulated halos from a quadratic Lagrangian bias model, working at relatively large scales in Section~\ref{sec:bias_results} and quantify the in-principle-accessible $\fnl$ information at the field level (at fixed phases) with this model using perturbative mocks in Section~\ref{sec:results_fnl}. We discuss the implications of these results in Section~\ref{sec:conclusions}.

\section{Field Level Local Primordial Non-Gaussian Bias Model}
\label{sec:model}

\subsection{LPNG bias}
\label{sec:lpng_bias_review}
Let us briefly review the LPNG bias signal in perturbation theory.
Local PNG is a simple phenomenological model for the Bardeen potential $\phi(\xv)$ such that
\begin{equation}
    \label{eqn:pheno_lpng}
    \phi(\xv) = \phi_{G}(\xv) + \fnlloc \left(\phi_{G}^{2}(\xv) - \langle \phi_{G}^{2}(\xv) \rangle \right),
\end{equation}
where $\phi_{G}$ is a Gaussian field\footnote{We neglect higher-order Local PNG, e.g. due to $g_{\mathrm{NL}},\tau_{\mathrm{NL}}$. See, e.g., Refs.~\cite{simone_gnl_taunl,kendrick_gnl_lpng_bias} for discussion.}.
Non-zero $\fnlloc$ in this model generates a modulation of short-scale physics by long-wavelength modes (e.g. generating a squeezed bispectrum, collapsed trispectrum, non-trivial position-dependent power spectrum, etc.).
Through a simple peak-background split argument \cite{Slosar08,DJS}, one can show that the local variance of scalar fluctuations is rescaled by the presence of a long-wavelength mode.
For a Fourier mode in the infinite-wavelength limit, $k\to 0$, this amounts to a rescaling of the amplitude of scalar fluctuations with finite wavelengths (i.e. $A_{s}$ or $\sigma_8$).
Since the number density of tracers (e.g. halos, galaxies) $n_{t}$ is sensitive to the amplitude of these fluctuations this produces a scalar bias
\begin{align}
    b_{\phi}^{(\mathrm{SU})} & = 2\frac{d \log
    n_{t}}{d \log \sigma_{8}}
    \label{eqn:su}
\end{align}
which captures the response of tracer number density to a change in the amplitude of fluctuations. We term equation~\ref{eqn:su} the variance Separate Universe (SU) \cite{baldauf_su2} version of LPNG bias to distinguish it from our field-level measurements. Assuming that the number density of tracers is a universal function of halo peak height \cite{Slosar08}
\begin{equation}
    \label{eqn:umf}
    b_{\phi}^{(\mathrm{UMF})} = 2 \delta_{c}b_{\delta},
\end{equation}
where $b_{\delta}$ is the linear Lagrangian bias, and $\delta_c=1.686$ is the critical overdensity threshold for spherical collapse.
This final approximation, or a slight modification of it, has frequently been used in the literature to model LPNG bias.

The above argument shows that, in the presence of LPNG, an extra term is needed in the bias expansion for tracer overdensity \cite{pat_bphi_08,DJS}. At leading order, we have in the  Eulerian bias expansion:
\begin{equation}
    \delta_{t}(\xv) = b_{1} \delta(\xv) + \bphi \fnl \phi(\xv) + \epsilon(\xv),
\end{equation}
where $\epsilon$ is a stochastic field. Terms involving $\bphi \fnl$ then enter the $n$-point statistics of the tracer field.
For example, the tracer power spectrum using only a linear (Gaussian) bias is augmented by a contribution that can be written as:
\begin{equation}
    \label{eqn:lpng_pk}
    P_{tt}(k) = (b_{1} + b_{\phi} \fnl \mathcal{M}^{-1}(k))^{2}P_{L}(k) + \frac{1}{\bar{n}},
\end{equation}
where $b_{1}$ is the Eulerian linear bias, $P_{L}(k) = \langle \delta(\kv) \delta(\kv') \rangle'$ for the linear matter overdensity field $\delta$, and $\mathcal{M}(k) = \frac{\delta(k)}{\phi(k)}$ 
is the normalized transfer function. The second term accounts for Poisson shot noise, which is typically subdominant on the linear scales on which this expression is good approximation.

\subsection{Field-level bias model}
\label{sec:bias_model_field}

In this work, we will go beyond linear bias by using a Lagrangian biasing model 
\cite{Matsubara08,Vlah16,ab18,Modi19} and nonlinear displacements obtained from N-body simulations (see Section~\ref{sec:n-body}) as in Refs.~\cite{chirag_sim_symm,kokron_stoch_heft,pellejero_hybrid_bias,pellejero_nbody_lagrangian_1,boryana_heft,kokron_heft_1,baradaran_heft,pellejero_2024_2,maion_heft_24}.
In detail, this means the Lagrangian fields that constitute the biasing operators used to model $\delta_{t}(\xv)$ are connected to their Eulerian positions by ``advection'' via the non-linear displacements, and are denoted by $\mathcal{O}^{o,\mathrm{adv}}(\qv)$.
As described in Ref.~\cite{chirag_sim_symm}, this is achieved by weighting the N-body matter tracer particles by the Lagrangian operators via gridded interpolation, and displacing them using the tracer particle locations at the final redshift of interest (see Section~\ref{sec:n-body} for numerical details).

We employ a 2nd-order Lagrangian bias model for biased tracers at the field level
\begin{align}
       \label{eqn:field_model_g}
        \delta_{t,\mathrm{fwd}}^{G}(\xv) &=   \delta(\xv) + b_{\delta} ~D(z)~\delta^{\rm{adv}}(\qv) \\
        &\quad \nonumber +b_{\delta^{2}}~ D^2(z)~\delta^{2,\rm{adv}}(\qv) + b_{K^2} ~D^2(z)~K^{2,\rm{adv}}_{ij}(\qv) \\
        &\quad \nonumber  + b_{\nabla^2 \delta}~D(z)~\left(\nabla^2 \delta\right)^{\rm{adv}}(\qv) \\
        &\quad \nonumber + \epsilon_{t}(\xv),
\end{align}
as well as its linear restriction, where $\delta(\xv)$ is the Eulerian matter density field, $D(z)$ is the linear scale-independent growth factor of matter fluctuations 
normalized to its $z=0$ value, $\epsilon_{t}$ is the tracer stochasticity field, and we include operators up to second order in $\delta$ as well as at leading order in derivatives. 
We note that the superscript ``$^{G}$'' indicates the usual ``Gaussian'' bias expansion, but, to be clear, all operators for $\fnl\neq0$ cosmologies are generated using an overdensity $\delta = \mathcal{M}\phi$ where $\phi$ is the \textit{non-Gaussian} field. 
Therefore these terms lead to, e.g., the leading primordial bispectrum contribution.
We note that, while we will mostly be concerned with halos here, this language is fully general for other biased tracers like galaxies.
The leading derivative correction partially absorbs numerical effects from the grid, due to both the finite resolution of the observed fields and missing dynamical contributions from short-wavelength modes, removing the leading-order resolution dependence of the model other than those already absorbed by the stochastic noise term $\epsilon_t$.
We neglect the quadratic stochasticity $[\epsilon_{\delta}\delta]$, and therefore the stochastic contributions to $b_{\delta}$, though we note that including it would tend to decrease the amount of PNG information at the field level especially in the more realistic case that the phases of the initial conditions are unknown.\footnote{The modification to the likelihood in Equation~\ref{eqn:eft_likelihood} including such density-dependent stochasticity was derived in Ref.~\cite{giovanni_stoch_eft_like}. Roughly, the noise $\sigma_0^2$ gains linear and quadratic dependencies on $\delta$, i.e. $\sigma_0^2 \rightarrow \sigma^2(\delta) = \sigma_0^2 + \sigma_1^2 \delta + \sigma_2^2 \delta^2$. In the fixed initial conditions setup that we employ throughout this work, these additional free coefficients can be, in principle, solved for numerically, though the process will be slightly more involved than for a constant $\sigma_0$ due to the density dependence in $\sigma(\delta)$ up- and down-weighting over(under)dense regions.}

The presence of LPNG introduces additional contributions to the bias expansion. To take these into account we augment the bias expansion by
\begin{align}
   \label{eqn:field_model_ng}
   \delta_{t,\mathrm{fwd}}(\xv)  = \delta_{t,\mathrm{fwd}}^{G}(\xv) &+ b_{\phi} f_{NL}^{\mathrm{loc}}  \phi^{\rm{adv}}(\qv) + b_{\delta\phi} f_{NL}^{\mathrm{loc}} D(z) \left[\phi \delta\right]^{\rm{adv}}(\qv),
\end{align}
where the quantity $\left[\phi \delta\right]^{\rm{adv}}$ is computed prior to smoothing (and is \textit{not} the product of the smoothed fields $\phi^{\rm{adv}}$ and $\delta^{\rm{adv}}$). 
We keep the above expansion to first order in $\fnlloc$ (see Ref.~\cite{Cabass22BOSS} for an estimate of the importance of these effects when measuring $\fnlloc$ from the power spectrum). The leading higher-derivative correction to  $b_\phi$ is approximately degenerate with $b_{\delta}$, especially at large scales, so we will not fit it separately. As in the Gaussian case above we neglect the second-order stochastic fields that PNG induces \cite{DJS}. 

Integrating over the stochastic field $\epsilon_t$, we use a real-space likelihood \cite{giovanni_stoch_eft_like,fabian_s8_eft,kokron_stoch_heft}:
\begin{equation}
    -2\log(L(\delta_{t}(\xv),\mathbf{p}) = \sum_{i=1}^{N_{g}^{3}} \left(2\pi \sigma_{0}^{2}\right) +  \frac{\left(\delta_t(\xv_i) - \delta_{t,\mathrm{fwd}}(\xv_i,\mathbf{p})\right)^2}{\sigma_{0}^2},
    \label{eqn:eft_likelihood}
\end{equation}
where the free parameters of the theory model are $\mathbf{p} = \{\fnlloc,b_{\delta},b_{\phi},b_{\delta^{2}},b_{K^{2}},b_{\nabla^{2}\delta},b_{\delta\phi}\}$ and we neglect the position-dependence of the noise ($\sigma_{0}^{2} = P^{0}_{\epsilon_t} 
 / V_{\mathrm{cell}}$, $V_{\mathrm{cell}} = \left(L/N_{g}\right)^3$). Since we assume the noise to be Gaussian, the inference of the biasing parameters reduces to simple linear regression with the best-fit bias parameters independent of $\sigma_0$.
We maximize the log likelihood to obtain bias parameters which implies that the best-fit $\sigma_{0}$ is given by the standard deviation of the residuals for the best-fit model.

\subsection{Numerical implementation}
\label{sec:n-body}

We evolve the matter field from redshift $z_{\rm{ini}} = 99$ to $z=0$ using \texttt{FastPM} \cite{fastpm} in a box of size $L = 2000 ~ \M$, with $N_{p}=1024^3$ simulation particles and force resolution ``boost'' factor of $B = 2$, using 40 timesteps.
We compute the linear theory power spectrum used to generate initial conditions using \texttt{CAMB} \cite{camb}, and use the $\Lambda$CDM parameters $\Omega_m = 0.3175$, $\Omega_b = 0.049$, $h = 0.6711$, $A_s = 1.91 \times 10^{-9}$ ($\sigma_8 = 0.834$) and  $n_s = 0.9624$.
We find halos using the \texttt{FastPM} implementation of Friends-of-Friends (FoF) with linking length $\ell=0.2$.
We consider a variety of halo mass bins, which are specified as necessary in subsequent sections.
We use simulations with LPNG amplitudes of $\fnlloc = \{ -100, -50, -25, -10, -5, -1, 0, 1, 5, 7, 10, 25, 50, 100 \}$.
We will work at redshift $z=1$ unless otherwise stated.

Injecting local PNG into the Gaussian initial conditions $\phi_G(\vk)$ is possible via the numerical implementation of equation~\ref{eqn:pheno_lpng} using real Discrete Fourier Transforms (RFFTs)
\begin{align*}
    \phi(\xv) =  \mathrm{RFFT}\left[\phi_{G}(\vk) + \fnlloc~\mathrm{RFFT}^{-1}\left[\mathrm{RFFT}[\phi_G]^{2}(\xv) - \langle \mathrm{RFFT}[\phi_G]^{2}(\xv) \rangle \right]\right].
\end{align*}
These local PNG initial conditions are implemented in the current publicly-available version of \texttt{FastPM} \cite{fastpm}, where the initial field $\phi_{G}(\kv)$ is smoothed with a sharp-$k$ filter beyond $k_{s} = \frac14 k_{\mathrm{Nyq}}$ to suppress numerical artefacts due to aliasing in the quadratic local transformation. We additionally checked for artefacts due to squaring the field on the grid by confirming that doubling the resolution of the initial conditions has no significant impact on our results.

We follow a procedure similar to that of Ref.~\cite{chirag_sim_symm} to compute the advected fields using \texttt{nbodykit} \cite{nbodykit}.
We compute Lagrangian operator fields $\delta,\delta^{2},K_{ij}^{2},\nabla^{2}\delta, \phi, \phi\delta$ from the linear matter overdensity field at the same resolution  $\Lambda$ as the initial conditions. 
We use LPNG-renormalized, or normal-ordered, quadratic Gaussian bias operators (see Appendix~\ref{app:renorm} for further details).
We deposit these fields to particles and match the particle IDs to those at their final (Eulerian) location at $z=1.0$ to advect the operators.
We then paint the advected particles according to their Lagrangian field weight to obtain the fields on grids of various resolutions $N_{g}^{3}$ to obtain the components of the real-space likelihood in Equation~\ref{eqn:eft_likelihood}. 
We will often also refer to the grid (Nyquist) wavenumber $\Lambda$ associated with a particular resolution $N_{g}$.
We compensate the linear mesh following Ref.~\cite{jing_compensation}.
When using linear regression to obtain the bias parameters or when estimating $\fnl$ in our profile likelihood procedure, we will always fit in real space using all modes on the grid with cell-size $\Delta = L/N_{g}$ (i.e., roughly, $r_{\rm{min}} \approx \Delta$, or $k_{\rm{max}} \approx \Lambda$; see Appendix~\ref{app:renorm} for further discussion).

\section{LPNG bias parameters}
\label{sec:bias_results}

In this section we present the results of our field-level bias measurements, including the LPNG bias parameter $\bphi$ as a function of the linear bias along with other bias parameters in the quadratic model. We discuss the implications for the field level information content with respect to $\fnlloc$ in the next section. 

\subsection{Universality at the field level}
\label{subsec:results_bphi_b1}

\begin{figure}[h!]
    \label{fig:bphi_b1}
    \centering
    \includegraphics[width=\textwidth]{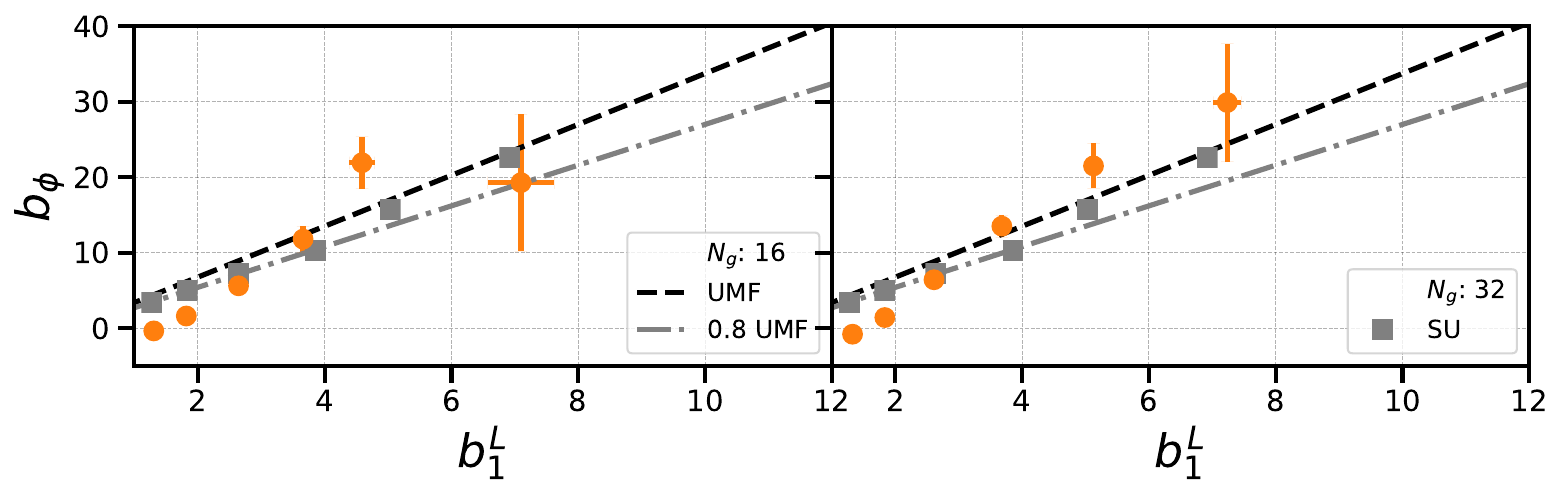} %
    \caption{
    \textit{Linear/PNG bias}
    The $\bphi(b_{1}^{L})$ measurements from field level fits for two (small) choices of cutoff wavenumber $\Lambda$, corresponding to the grid scale with $N_{g}=\{16,32\}$ 1-dimensional cells in the left and right panels, respectively.
    The bias values measured with the field-level model (and their associated uncertainties) are shown in orange.
    The separate universe measurements are plotted as gray squares, and the universality relation is plotted in black dashed, with its empirical modification (by a factor of 0.8) plotted in gray dashed.}
\end{figure}

Figure~\ref{fig:bphi_b1} shows the measured values of $\bphi$ and $b_1^{L}=b_{\delta}$ using the field-level model with Gaussian errorbars and linear regression for six mass bins that are equally spaced in log halo mass $\log\left(\frac{M_{h}}{M_{\odot}/h}\right)$. The results here were obtained using a linear bias model varying only $b_\delta,\bphi$, but using a quadratic model yields quantitatively similar results.
The squares show the variance separate universe predictions for $\bphi$ plotted against $b_{1}^{L}$ values that are obtained by fits to power spectra.
In detail, we obtain the linear bias (used for the SU points in Figure~\ref{fig:bphi_b1}) as the large-scale ratio of the halo-matter cross power spectrum $P_{hm}(k)$ to the matter power spectrum $P_{mm}(k)$ up to $k_{\mathrm{max}}=0.05~h/\mathrm{Mpc}$.
For the values of the variance SU bias, we measure $b_{\phi}^{(\mathrm{SU})}$ using the version of eqn.~\ref{eqn:su} as in Ref.~\cite{2020JCAP...12..013B} (centered finite difference in relative amplitude change $\delta \sigma_8$)
at $\sigma_{8} = 0.834(1\pm \delta\sigma_8)$, with $\delta \sigma_8=0.02$. Using a step in $\sigma_8$ twice or half as wide has sub-percent impact on the estimated bias values.

The black dashed line shows the universal mass function (UMF) prediction $b_{\phi}^{(UMF)} = 2\delta_{c} b_{\delta}$ of eqn.~\ref{eqn:umf}.
The gray dash-dotted line shows $b_{\phi} = 0.8 \times 2\delta_{c}b_{\delta}$, which has been shown to provide a better fit to dark matter halos in simulations \cite{2020JCAP...12..013B}.
Overall, we see that the the field-level fits qualitatively agree with the UMF and SU predictions within the errorbars for large values of the bias.
However, at small values of the linear bias (small halo masses) we see some multi-$\sigma$ disagreement - we have verified this is not due to several aspects of our simulation resolution 
or the approximate nature of FastPM - see Appendix~\ref{app:sims} for further discussion.

\subsection{Field-level bias measurements at quadratic order}
\label{subsec:results_quad_bias}

\begin{figure}[h!]
    \centering
    \includegraphics[width=\textwidth]{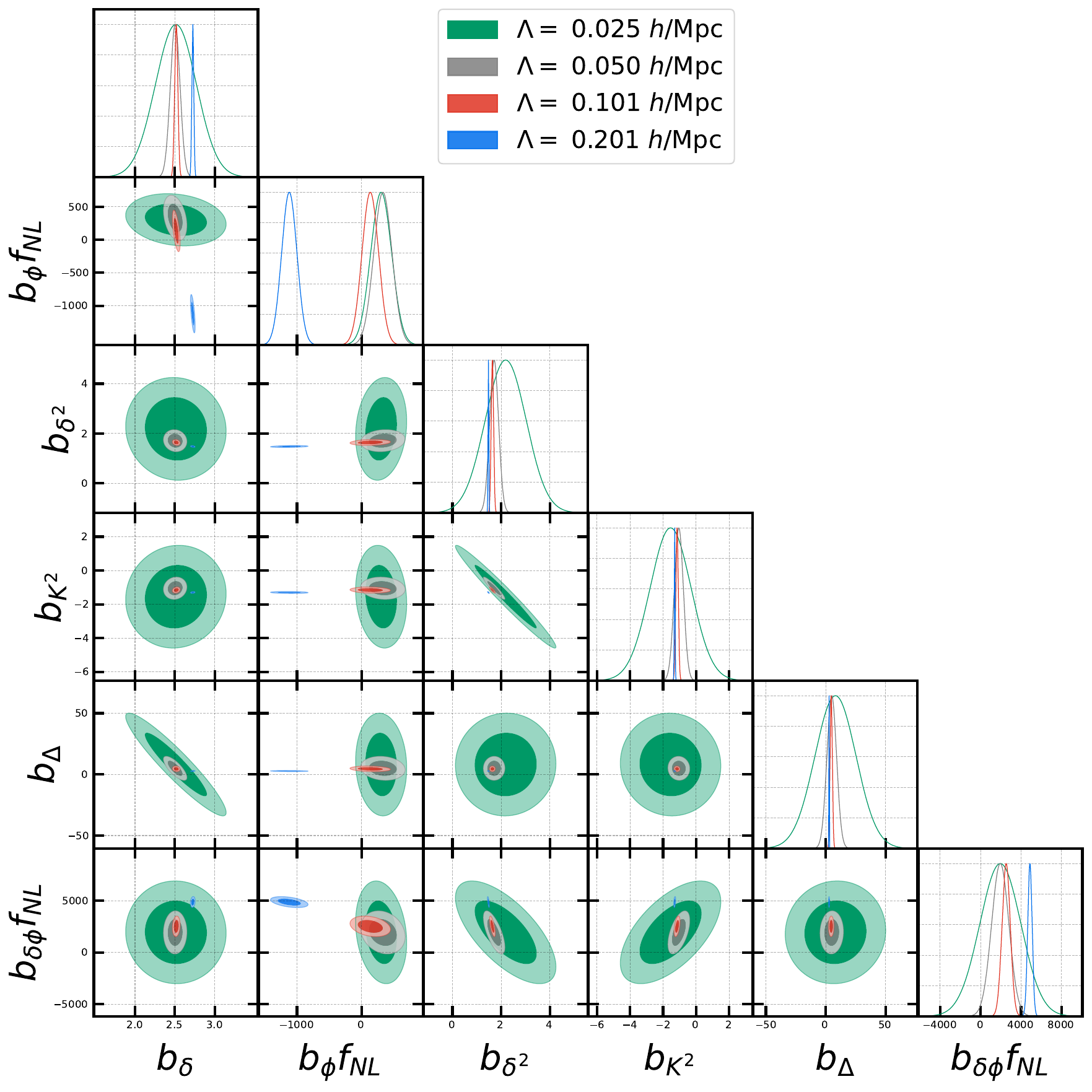} %
    \caption{\textit{Quadratic/PNG bias values}: Constraints on quadratic Local PNG bias parameters obtained from linear regression at several grid smoothing scales $\Lambda$ (colors) for a mass bin, $\log_{10}\left(\frac{M_{h}}{M_{\odot}/h}\right) = [13.62,13.71]$.}
    \label{fig:bphi_contour}
\end{figure}

Figure~\ref{fig:bphi_contour} shows field-level constraints on the quadratic and LPNG Lagrangian bias parameters when including the nonlinear bias terms in Equations~\ref{eqn:field_model_g} and \ref{eqn:field_model_ng}.
These Gaussian contours are obtained through simple linear regression using the field-level likelihood in eqn.~\ref{eqn:eft_likelihood}, which is equivalent to considering all real-space 2-point functions of the linear and quadratic operators. We obtain nearly identical results---up to differences in scale cuts---working in Fourier space.
The data vector consists of a mass-binned $N$-body halo density field obtained from a simulation with $\fnl=100$.
The likelihood noise $\sigma_0$ is set to its optimal value (estimated from the residuals in eqn.~\ref{eqn:resid}) and no priors are placed on the bias parameters.
We note that to obtain sensible constraints from this quadratic bias model, it is crucial to use the renormalized quadratic Gaussian component fields $(\delta^{2}, K^{2})$ as part of the regression basis, a point we discuss further in Appendix~\ref{app:renorm}.

The constraints on the linear LPNG bias, in the combination $\bphifnl$, are shown in the second column.
Considering first the 1D posterior at the top, it is apparent that the constraining power on $\bphifnl$ does \textit{not} change much with the likelihood cutoff scale $\Lambda$.
On the other hand, the constraints on the Gaussian bias parameters in Equation~\ref{eqn:field_model_g} noticeably improve as $\Lambda$ increases. This is as expected, since the information on $\bphifnl$ (at fixed $\fnl$) is coming from large scales. Importantly, we do not see much correlation between the linear LPNG bias with the other bias parameters, while all the quadratic bias parameters show nontrivial degrees of correlation, including the LPNG one $b_{\delta \phi} f_{NL}^{\rm loc}$.

The fact that the contours are nested within each other at the 1-$\sigma$ level for all scales $\Lambda$ up until the maximum value of $\Lambda=0.2 ~h/\mathrm{Mpc}$ indicates that the model is consistently fitting the bias parameters across scales (i.e. there is very limited ``running'' of the bias parameters with $\Lambda$).
The exception to this statement is the $\Lambda=0.2 ~h/\mathrm{Mpc}$ (blue) contour, where a systematic bias is produced\footnote{We verified that this bias at maximum $\Lambda$ is reproduced in fits to Gaussian simulations as well, indicating the bias is not due to the inclusion of PNG bias.}, which we now discuss.

\begin{figure}[h!]
    \centering
    \includegraphics[width=\textwidth]{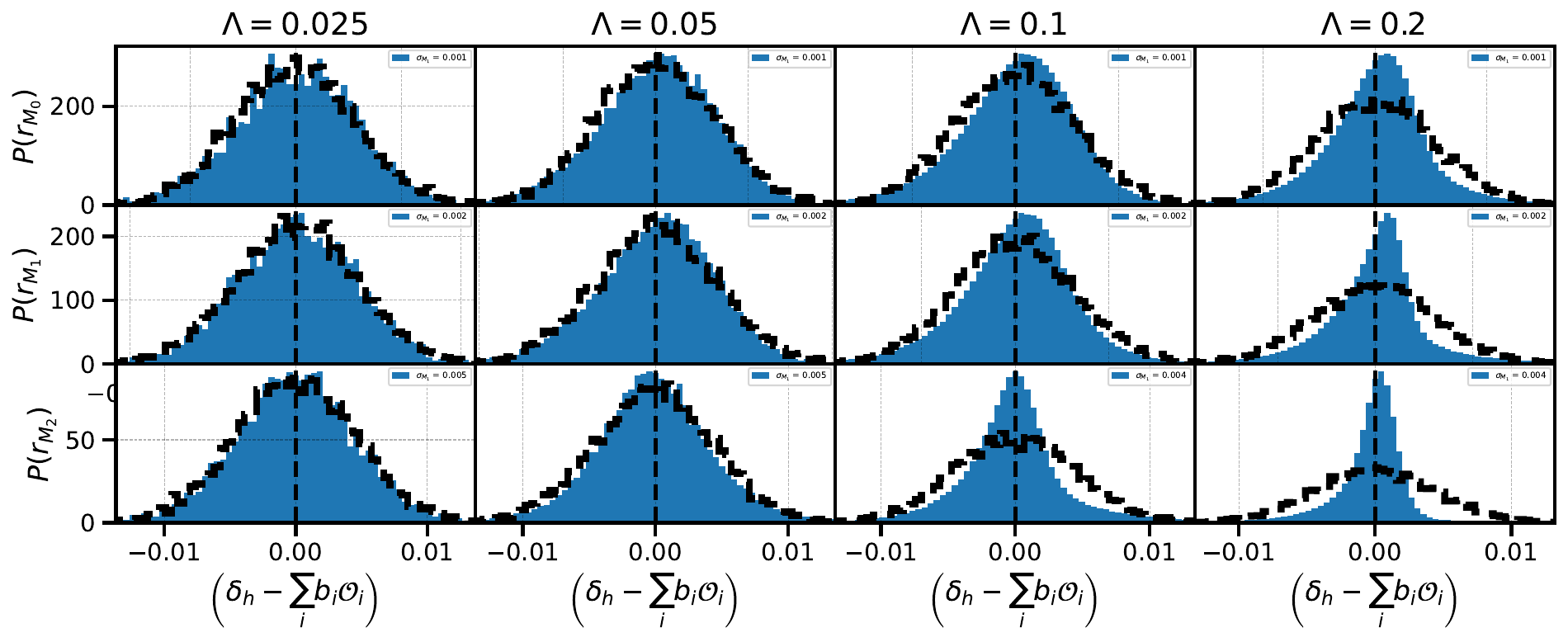} %
    \caption{\textit{Residual PDFs}: Residual probability density functions for several choices of halo mass and smoothing scale using the quadratic biasing model. The three halo mass bins shown have bin edges $\log_{10}\left(\frac{M_{h}}{M_{\odot}/h}\right) = \{13.1,  13.5, 14.0, 14.5\}$, and increase with increasing row number, while the grid cutoff scale ($\Lambda$, in $h/\mathrm{Mpc}$) used for the fields increases with increasing column number. For visual clarity, the residuals are normalized by the number of grid volume elements, $N_{g}^{3}$.}
    \label{fig:bphi_pdf}
\end{figure}

To assess the reason for the large bias at the largest value of $\Lambda$, we can consider the probability density function (PDF) of the field-level residuals entering the likelihood
\begin{equation}
    \label{eqn:resid}
    r_{M}(\mathbf{x}) = \delta_{h,M}(\mathbf{x})  -\delta(\mathbf{x}) - \sum_{i} b^M_{i}\mathcal{O}^{(i)}(\mathbf{x}).
\end{equation}
Figure~\ref{fig:bphi_pdf} shows this PDF for all halo mass bins with variation in smoothing scale and $\fnlloc$ value.
The empirical scatter $\sigma_{0}$ is quoted in the legend for each PDF.
As expected, going to higher halo masses (down the column) produces increased values of $\sigma_{0}$, as there are fewer high-mass halos.
Similarly, we see that going to higher values of the cutoff $\Lambda$ results in more non-Gaussian densities - this behavior has been seen previously for field-level bias modeling (see e.g. Table III of Ref.~\cite{marcel_halo_field}). 
Already some non-Gaussianity in the PDF is developing for the most massive halos at $\Lambda = 0.1 ~h/\mathrm{Mpc}$, though this appears not to affect the bias parameter constraints obtained from the likelihood fits in Fig.~\ref{fig:bphi_contour}.
Clearly at the largest value of $\Lambda$ (corresponding to $N_{g}=128$) our Gaussian field-level likelihood breaks down significantly, and we do not expect it to produce unbiased constraints on $\bphi \fnl$, and we will see that this is indeed the case in Section~\ref{subsec:fnl_info_halos}.

\section{Field-level $\fnlloc$ information}
\label{sec:results_fnl}

\begin{figure}[h!]
    \centering
    \includegraphics[width=\textwidth]{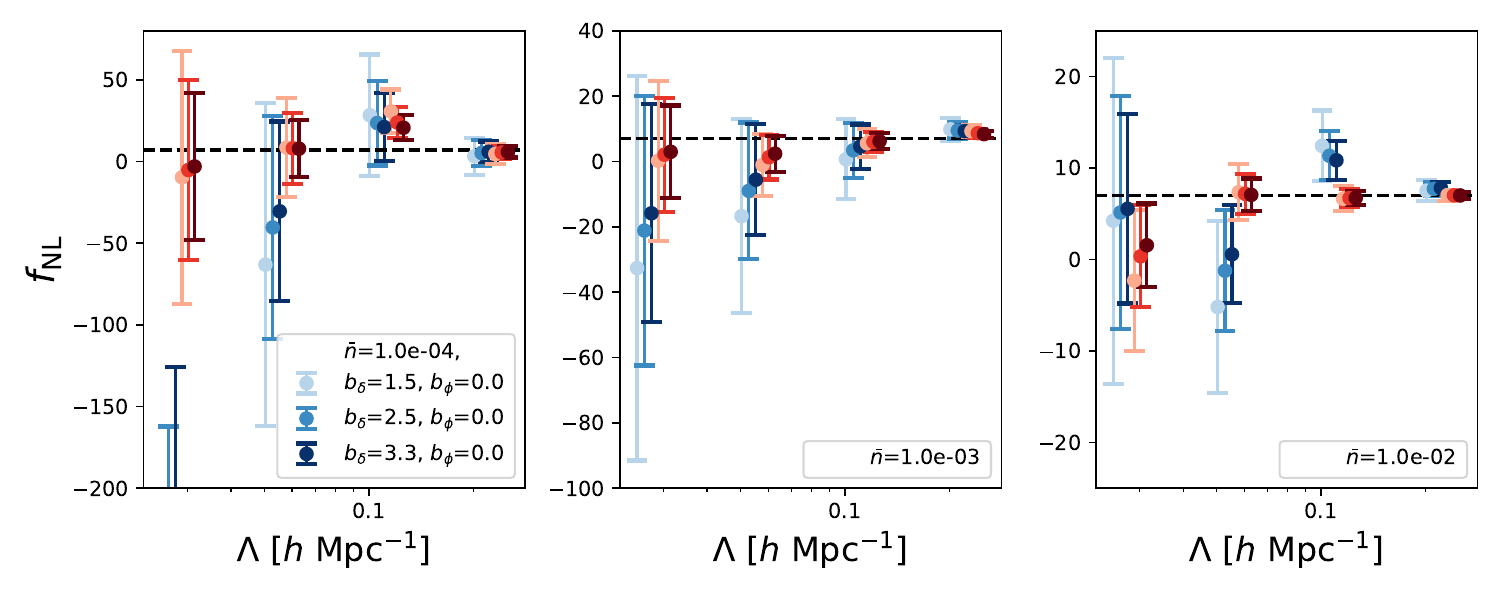}
    \caption{\textit{Zero quadratic/PNG bias}: Field-level constraints on $\fnlloc$ obtained from perturbative mocks.
    Values of $\fnlloc$, $\sigma(\fnlloc)$ are obtained from fitting the profile likelihood curvature and are shown as a function of the cutoff scale $\Lambda$.
    The true value of $\fnl=7$ used to generate the mock is marked by the dashed black line.
    Red and blue colors correspond to constraints obtained from the linear and quadratic cases.
    Different shades of a single color correspond to different input linear biases for the mocks.
    Each panel corresponds to a different level of Poisson noise injected into the mock.
    Constraints from mocks with distinct bias values have been horizontally offset for clarity.
    (We draw the reader's attention to the changing vertical axis in each panel.)}
    \label{fig:mock_pl_zerotest}
\end{figure}

\begin{figure}[h!]
    \centering
    \includegraphics[width=\textwidth]{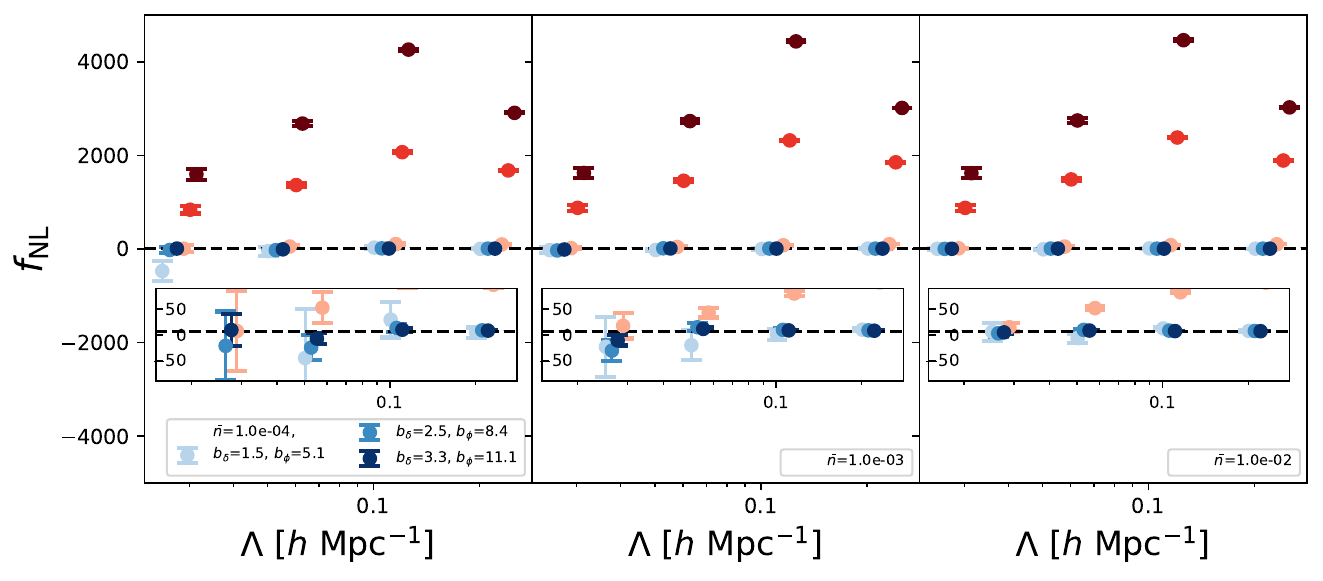} %
    \caption{\textit{Non-zero quadratic/PNG bias}: Similar to Figure~\ref{fig:mock_pl_zerotest}, but for nonzero values of the quadratic and PNG bias parameters.
    The inset shows a significant zoom-in of the y-axis, and reveals the unbiasedness of the quadratic model (blue points) relative to the linear model (red points).}
    \label{fig:mock_pl_fit}
\end{figure}

To test the potential to constrain $\fnlloc$ at the field level, we infer $\fnlloc$ by way of a profile likelihood procedure.
Rather than attempt to marginalize over initial conditions, we work with fixed initial conditions.
This provides a clear picture of the constraining power of the field level model in an idealized case, providing an upper bound on the information content of the field using our model.
These results, of course, should not be interpreted as reflecting what uncertainty on $\fnl$ is actually achievable for real LSS tracers, but rather are meant to quantify the information on $\fnl$ that is available at the field level in different scenarios when marginalizing over bias parameters. 
In practice, the initial conditions must be marginalized over in order to constrain $\fnlloc$ in a full field-level inference framework - a task that is an area of vigorous ongoing research \cite{adrian_hmc,adrian_hmc_2,marius_lensing_22,uros_field_level,pmwd,nguyen_field_level,biwei_multiscale_flow,feng_recon,borg_1,chen_cnn,chen_ml_png,cabass_field_level,modi_sbi,lemos_field_level,ramannah_hmc,floss_nn_24,jamieson_emulator_2}.
We consider the case of both perturbative mocks (generated by adding noise to our bias model) and simulated dark matter halos in this Section.

\subsection{Perturbative mocks}
\label{subsec:fnl_info_mock}
We perform two tests of $f_{NL}$ information contained in the field using perturbative mock tracers generated from our bias model.
Figures~\ref{fig:mock_pl_zerotest} and \ref{fig:mock_pl_fit} show constraints from a field-level profile likelihood procedure, similar to that used by Ref.~\cite{fabian_s8_eft} to constrain $\sigma_8$ from simulated halos at fixed initial conditions\footnote{We are able to approximately reproduce the results of Ref.~\cite{fabian_s8_eft} with our procedure, e.g. in their Fig.~3, though the results are not directly comparable due to differing forward models.}.
We set as our data a perturbative mock with a value of $\fnl=7$, and compute the maximum likelihood value of $\fnl$ for each of our $\fnl$ simulations, marginalizing over the bias parameters and fitting for $\sigma_0$.
By fitting a parabola to the log likelihood, we obtain the central value and $\pm1\sigma$ uncertainties on $\fnl$.
The bias parameters are implicitly marginalized over with uniform priors during linear regression (see Appendix~\ref{app:priors} for other choices of prior).
We perform this procedure for several smoothing scales $\Lambda=\{0.025,0.05,0.1,0.2\}h/\mathrm{Mpc}$ (horizontal axis).
Each panel of Fig.~\ref{fig:mock_pl_zerotest} and   Fig.~\ref{fig:mock_pl_fit} shows the constrained value of $\fnlloc$ from the profile likelihood procedure applied to a mock generated from the quadratic biasing model with a different level of injected Poisson noise, corresponding to number densities of $\bar{n} = \{10^{-4}, 10^{-3}, 10^{-2}\}~[h/\mathrm{Mpc}]^3$.
We use mock linear bias parameters of
$b_\delta=\{1.5,2.5,3.3\}$, which are represented by different color shades. The red points show results for $\fnlloc$ using a linear model and the blue points for using a quadratic model.
The lightest (darkest) shade of red and lightest (darkest) shade of blue correspond to a mock with the same linear bias value fitted by the linear model and quadratic model, respectively.

In Figure~\ref{fig:mock_pl_zerotest} we set all bias parameters to zero except for $b_\delta$, including the PNG bias coefficients.
This test therefore probes the impact of simply varying the quadratic bias parameters rather than fixing them on the $\fnlloc$ information contained in the density field at the field level.
In each panel, we see that increasing the cutoff scale $\Lambda$ results in increasingly tight constraints on $\fnl$ when marginalizing over bias parameters.
This indicates that we are accessing nonlinear information contained in the density field. 
However, when comparing the linear (red) and quadratic (blue) mock and model results, we see that in each panel, for any choice of mock bias parameters (same shade) or cutoff scale $\Lambda$, the errorbars increase by a factor of more than 2-3.
The non-linear information being probed is apparently obscured by the lack of knowledge of the quadratic bias parameters. This is consistent with existing results in the literature \cite{2022JCAP...11..013B,Cabass22BOSS}, which suggest that, at least for volumes of existing survey data, the primordial contribution to the tracer bispectrum does not lead to improved constraints on $\fnl$ in practice.
In fact, here we can make an even stronger statement about the low impact of this term on $\fnl$ information since we work at the field level and at fixed phases.

In Figure~\ref{fig:mock_pl_fit} we set bias parameters to those fitted to halos (as a function of $b_\delta$) in the literature.
Specifically, $b_\phi$ is set with the universality relation (eqn.~\ref{eqn:umf}), $b_{\delta^2}$ and  $b_{K^2}$ are set using the (Gaussian) measurements of Ref.~\cite{ab18}, $b_{\nabla^{2}\delta}$ is fixed to zero, and $b_{\delta\phi}$ is fixed to its universality prediction\footnote{E.g. as stated in Ref.~\cite{2022JCAP...01..033B} eqn. 1.5.}.
As expected, in each panel at fixed number density, the larger the true value of $\bphi$, the larger the $\bphifnl$ signal at fixed $\fnlloc$, resulting in tighter constraints on $\fnlloc$\footnote{We also note that the amount of injected noise, which is separate from the bias value used to generate the mock in this context (as opposed to the case of halos/realistic tracers), generally appears not to bias the inferred value of $\fnlloc$.}.
While in Fig~\ref{fig:mock_pl_zerotest} the linear model could accurately fit the quadratic mock, here obviously this will not be the case.
However, it is quite striking that even on very large scales (low $\Lambda$) we see that the linear model (red points) generally returns extremely biased results when attempting to fit to a quadratic perturbative mock.
This is an indication that even on very large scales, it is not possible to constrain $\fnlloc$ for a nonlinear tracer like halos and galaxies without also marginalizing over quadratic bias parameters, i.e. significant degeneracies exist between higher-order bias parameters and $\fnlloc$ even at the field level when $b_\phi$ is free. Heuristically, without knowing the value of $b_\phi$, the leading signal for $\fnlloc$ comes primordial contributions to the bispectrum; however, since the leading contributions to the bispectrum are gravitational and depend on the values of $b_{\delta^2}, b_{K^2}$, fixing the values of the nonlinear bias parameters leads to a misattribution of the primordial signal and thus biased $\fnlloc$ constraints even for small $\Lambda$.

\subsection{Simulated halos}
\label{subsec:fnl_info_halos}

\begin{figure}[h!]
    \centering
    \includegraphics[width=\textwidth]{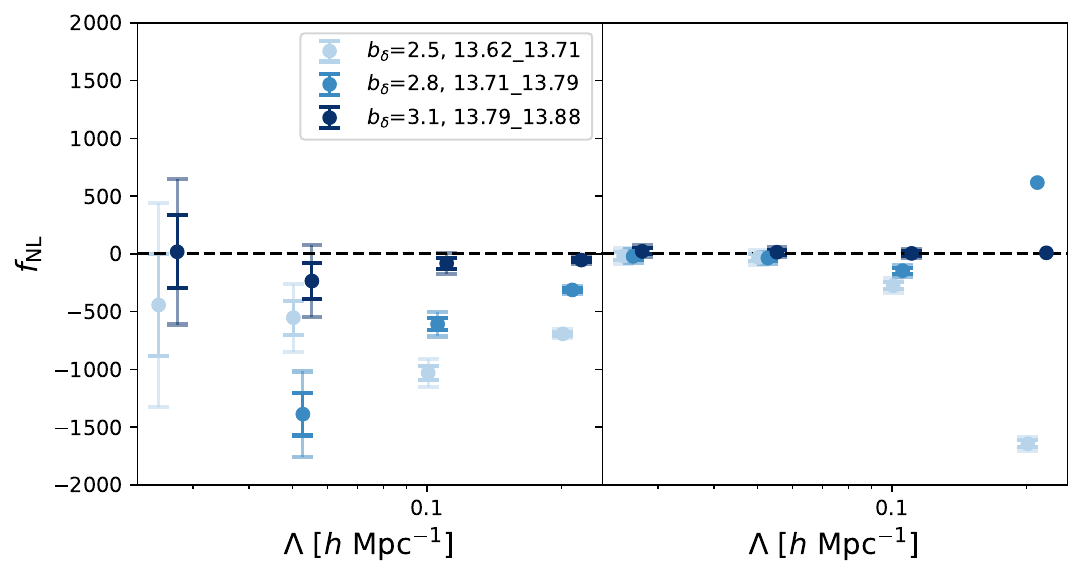} %
    \caption{\textit{Quadratic + halos:} Field-level constraints on $\fnlloc$ obtained from N-body halos using a quadratic biasing model (at 1-$\sigma$ and 2-$\sigma$).
    As in Figs.~\ref{fig:mock_pl_zerotest},~\ref{fig:mock_pl_fit}, constraints are obtained from our profile likelihood procedure.
    The true value of $\fnl=7$ used to run the simulation is marked by the dashed black line.
    \textit{Left:} Constraints from halos when marginalizing over all bias parameters. 
    It is clear that results are biased, especially for certain halo mass bins.
    \textit{Right:} The same profile likelihood procedure, but where all values of the PNG biases $b_{\phi}, b_{\delta\phi}$ are fixed to their best-fit values for \textit{each} $\fnlloc$, thus breaking the $\bphifnl$ degeneracy.
    We see that in this case, constraints are unbiased for the halo sample in question for all but the highest $\Lambda$.
    }
    \label{fig:halo_pl_fit}
\end{figure}

\begin{figure}[h!]
    \centering
    \includegraphics[width=\textwidth]{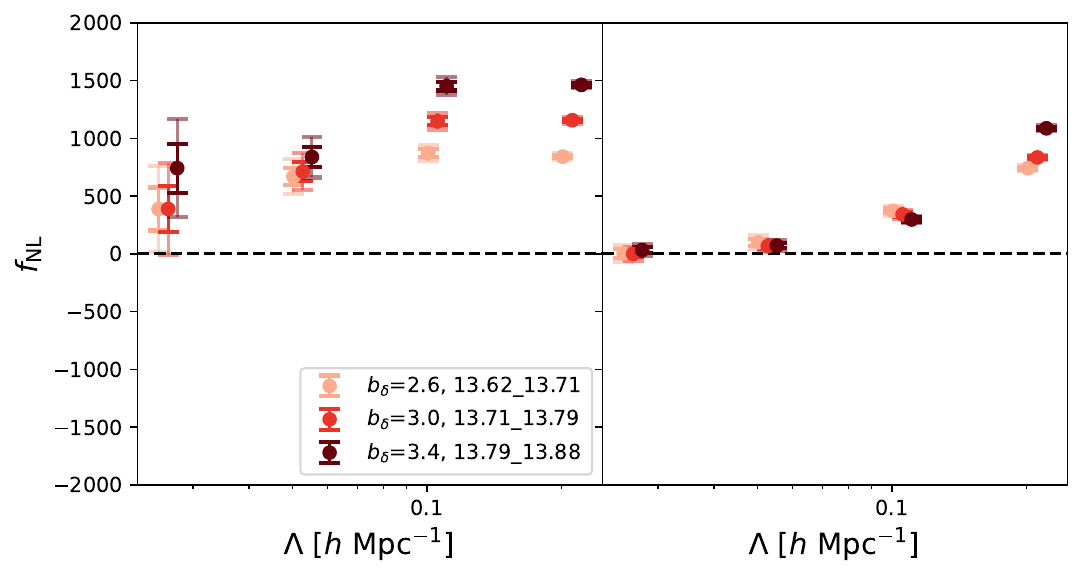} %
    \caption{\textit{Linear + halos:} Similar to Figure~\ref{fig:halo_pl_fit}, but using a linear bias model. 
    By comparison, the results are generally significantly more biased (with smaller error bars), even on large scales, and are biased toward high $\fnlloc$, while the opposite is true for the quadratic model in Fig.~\ref{fig:halo_pl_fit}.
    The right panel also shows the expected failure of the linear bias model when the PNG biases are fixed for all but the lowest values of $\Lambda$.
    }
    \label{fig:halo_pl_fit_lin}
\end{figure}

Figure~\ref{fig:halo_pl_fit} shows profile likelihood $\fnlloc$ constraints obtained from N-body halos with a quadratic biasing model.
In general, we find in the left panel that constraints on halos are significantly biased, though there are some mass bins for which this is not the case.
This is due to the $\bphifnl$ degeneracy\footnote{We thank Fabian Schmidt for productive discussions around this point.}:
the right panel of Fig.~\ref{fig:halo_pl_fit} shows what happens when we perform the profile likelihood procedure with PNG bias parameters fixed to their best fit values for each $\fnlloc$ simulation\footnote{We note that for the smaller values of $\fnlloc$, the value of $\bphi$ is not well constrained. The results are therefore most sensitive to the fitted values of the  simulations with larger $\fnlloc$.}, which completely breaks this degeneracy.
Except for the highest $\Lambda$, for which we showed the quadratic bias model is insufficiently accurate in Section~\ref{subsec:results_quad_bias}, the constraints are unbiased.
While for the mock test we showed that there is significant $\fnlloc$ information in the nonlinear density field (Figs.~\ref{fig:mock_pl_zerotest},\ref{fig:mock_pl_fit}), for N-body halos it appears that constraints obtained from fitting a quadratic model are biased due to the $\bphifnl$ degeneracy.
This suggests that marginalizing over the PNG bias parameters, and hence the large-scale local PNG bias signal, is not a feasible strategy for constraining $\fnlloc$ at the field level from halos.
In short, unbiased constraints of $\fnlloc$ from tracers at least as complicated as simulated halos using a quadratic bias model are only attainable with prior knowledge on $\bphi$ and $\bdphi$.
Figure~\ref{fig:halo_pl_fit_lin} shows that the difficulty is even more pronounced in the case of a linear bias model, and persists even at the lowest $\Lambda$ where the linear model should be accurate.
The errorbars on the halo constraints are relatively large compared to the mocks due to the narrow mass bins we consider for these simulations - it would be interesting to extend these results with larger volume simulations and higher number densities.

\section{Conclusions}
\label{sec:conclusions}

Local primordial non-Gaussianity (LPNG) is a signal of the inflationary era that can be probed with large-scale structure surveys.
Given recent technological and methodological developments, performing field-level inference of cosmological parameters using the galaxy distribution is becoming increasingly feasible.
In this paper, we use a field-level Lagrangian biasing model to perform the simple extension of existing perturbative field-level models to the case of LPNG at quadratic order while using fully non-linear displacements.
We use this machinery to test the widely-used universality relation between the LPNG bias $\bphi$ and the linear (Eulerian) bias $b_1=1+b_{\delta}$ for simulated dark matter halos, finding rough agreement with results from variance separate-universe simulations.
We also obtain constraints on quadratic bias parameters, including LPNG parameters $b_{\phi},~b_{\delta\phi}$ of simulated halos at the field level using several smoothing scales $\Lambda$, finding that the results from our biasing model are stable below wavenumbers of $\Lambda=0.2~h/\mathrm{Mpc}$ at $z=1$. 
Finally, we use perturbative mocks, simulated dark matter halos, and a field-level profile likelihood to obtain constraints on $\fnlloc$ while marginalizing over all bias parameters, including LPNG bias parameters---we find that using a quadratic bias model, as is necessary for all but the largest scales probed by cosmological observations, degrades the constraints on $\fnlloc$ by a factor of greater than 2-3.
We also find that constraints on $\fnlloc$ from halos are generally quite biased, even with the quadratic model, due to the $\bphifnl$ degeneracy.
We can make several conclusions based on these results.

Our simple test of the $f_{NL}$ information available at the field level when marginalizing over nonlinear bias has implications for recent efforts toward field-level inference of cosmological parameters, including $\fnlloc$.
We show that there \textit{is} information that can be captured from the field level with our bias model when marginalizing over all bias parameters.
Naively, this would suggest that we are able to extract significant nonlinear information from the tracer density field, which is in agreement with, e.g. Refs.~\cite{quijote_png, quijote_png_halo, quijote_png_halo_2,andrews_borg_png}, as we also found that constraints on the combination $\bphi f_{NL}$ (at fixed $\fnl$) are essentially unaffected by changes in $\Lambda$, so any additional information must come 
from the matter density field itself. 
However, we also show that using a quadratic bias model (as necessary for using quasi-linear scales), 
degrades these constraints by a factor of a few - indicating that the nonlinear information is significantly diluted by their degeneracy with quadratic bias parameters.
Furthermore, we found that the inference of $\fnlloc$ from perturbative mocks generated using realistic values of quadratic halo bias parameters (including PNG bias parameters) is significantly biased when using a linear biasing model, showing that $\fnlloc$ at the field level requires careful marginalization of nonlinear bias even at the largest scales.
Finally, we showed that the inference of $\fnlloc$ from simulated N-body halos when marginalizing over bias parameters (including PNG bias parameters) is in general quite biased due to the $\bphifnl$ degeneracy.
This suggests that trustworthy priors on PNG bias parameters will be essential for future constraints of $\fnlloc$ from data using the scale-dependent bias signal.

The tests performed here on simulated dark matter halos, while far from observations, are the first use of the field-level bias model with LPNG to model the bias $\bphi$.
This step is mechanically simple but will be a necessary piece of any future perturbative field-level inference of $\fnl$ from tracers of large-scale structure.
An obvious extension would be to further develop our LPNG extensions of the field-level biasing framework to redshift space \cite{marcel_rsd_field,stadler_eft_field_rsd,julia_field_level_galaxies_24}.
Another application would be to measure LPNG bias from expensive hydrodynamical simulations where it is infeasible to run multiple cosmologies required for separate universe bias estimates.
This is especially of interest if it is possible to link simulated tracers (galaxies, QSOs, line intensity, etc.) and particular target samples of a redshift survey.
It would also be interesting to extend this work to field-level biasing models of non-local types of PNG.

Our results indicate that constraints on LPNG from large-scale structure will depend on informative $\bphi$ priors.
Such priors may be accessible from simulations e.g. through simulation-based priors \cite{sullivan_21,ivanov_priors, misha_structure_eft_prior,zhang_eft_prior}, internal consistency relations such as the $b_2(b_1)$ relation, suitably extended beyond the case of dark matter halos, or alternative methods of estimating $\bphi$.
Here we operate at fixed phases, which is optimistic for higher order correlations, but multi-tracer analyses can remove the random effects of phases by canceling cosmic variance in $n$-point statistics where the LPNG bias contribution dominates \cite{sullivan_23,barreira_krause_23,karagiannis_bispectrum_multitracer,heinrich_spherex_bispectrum_24,fondi_lpng_mt_forecast,hamaus_halo_forecast,ginzburg_shotnoise_multitracer,seljak_mt}.
While there is significant $\fnlloc$ information in the nonlinear density field, it seems likely that constraints from large-scale structure will continue to be driven by the large-scale bias contribution.

\acknowledgments
We thank Kazuyuki Akitsu, Martin White, Matias Zaldarriaga, and especially Fabian Schmidt for useful discussions of field-level bias modeling, as well as Uro\v{s} Seljak and Simone Ferraro for helpful conversations and Biwei Dai for pointing out a bug in FastPM on NERSC's Perlmutter.
JMS also thanks members of the Quijote PNG group for discussions.

JMS was partially supported by the U.S. Department of Energy, Office of Science, Office of Advanced Scientific Computing Research, Department of Energy Computational Science Graduate Fellowship under Award Number DE-SC0019323 as well as a U.S. Dept. of Energy SCGSR award during the completion of this work. 
JMS also acknowledges that, in part, support for this work was provided by The Brinson Foundation through a Brinson Prize Fellowship grant.
SC acknowledges the support of the National Science Foundation at the Institute for Advanced Study through NSF/PHY 2207583.

This research used resources of the National Energy Research Scientific Computing Center (NERSC), a Department of Energy Office of Science User Facility using NERSC award HEP-ERCAP0028635.
This research has made use of NASA's Astrophysics Data System.

\appendix

\section{Simulation dependence}
\label{app:sims}

\begin{figure}[h!]
    \centering
    \includegraphics[width=0.32\textwidth]{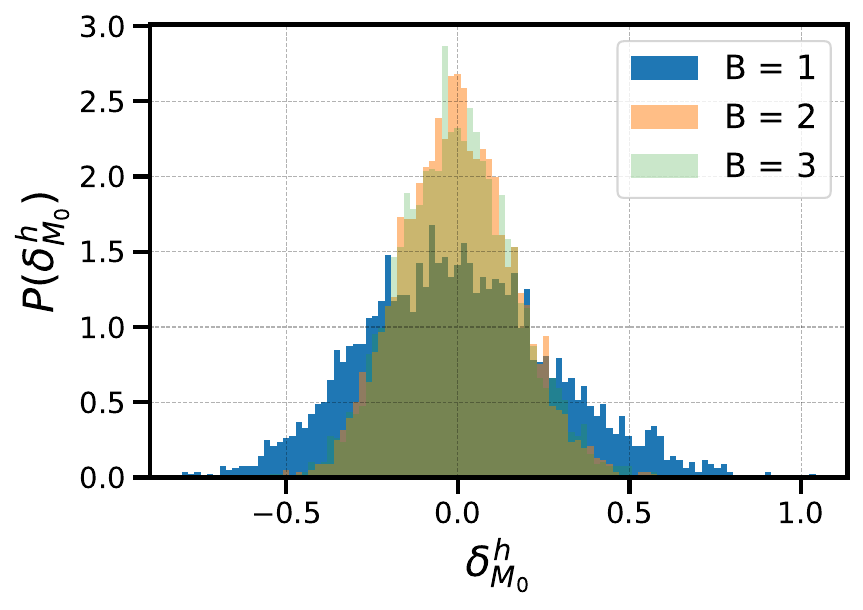} %
    \includegraphics[width=0.32\textwidth]{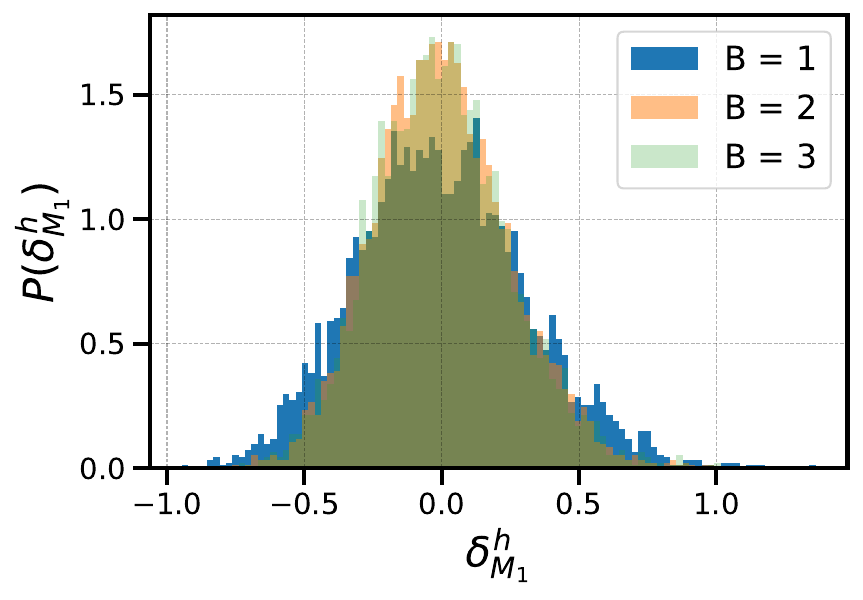} %
    \includegraphics[width=0.32\textwidth]{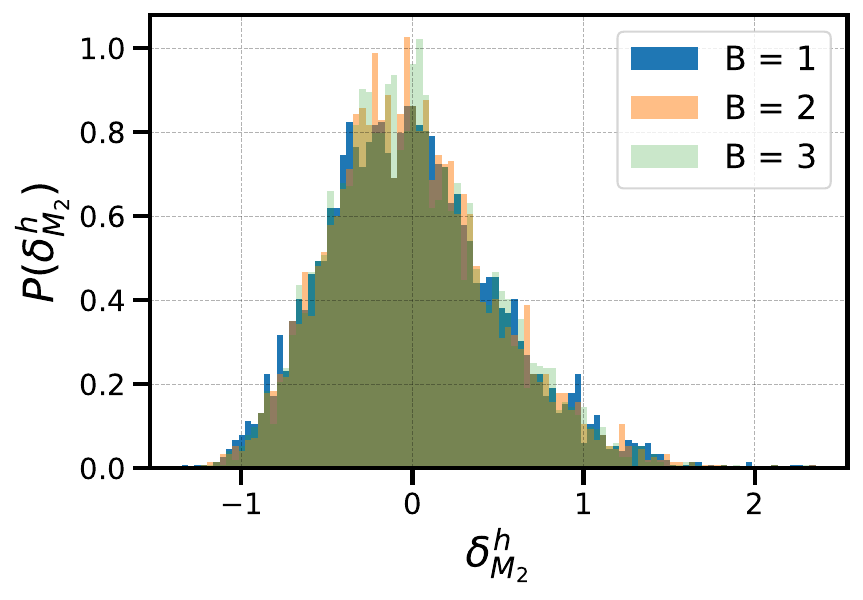} %
    \caption{Histograms of the halo density field for several choices of force resolution in FastPM with $\fnl=100$. There is a significant difference between the $B=1$ and $B=2$ cases, but not between $B=2$ and $B=3$.}
    \label{fig:app_fastpm}
\end{figure}

To test the extent to which our results depend on the resolution of the \texttt{FastPM} simulations used to advect the Lagrangian operators to their final positions, we test several force resolutions in FastPM.
Figure~\ref{fig:app_fastpm} shows histograms for three redshift bins (those shown in Fig.~\ref{fig:bphi_pdf}) where each color corresponds to a different $B$ corresponds to a different force resolution\footnote{The ``boost'' factor, $B$, is defined as the ratio of the force grid to the initial conditions grid \cite{fastpm}.}. 
Here we see that the $B=1$ case is significantly different from $B=2$ and $B=3$, the latter two of which agree well.
We use $B=2$ simulations in the main text.

\begin{figure}[h!]
    \centering
    \includegraphics[width=0.6\textwidth]{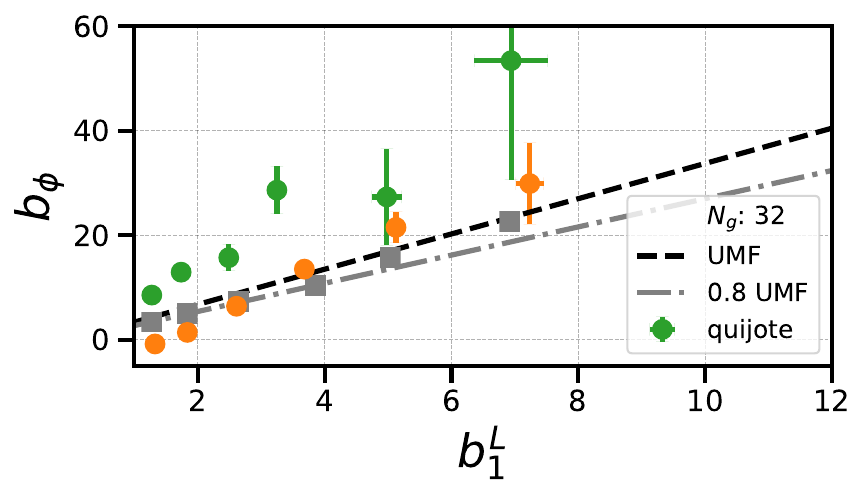} %
    \caption{Similar to Figure~\ref{fig:bphi_b1}, but with bias parameters measured for multiple FastPM 
    as well as the Quijote-PNG simulations (green).
    The bias measurements for Quijote are consistent with those of FastPM at high values of $b_{1}$ $\Lambda\approx0.05~h/\mathrm{Mpc}$, but for low values of $b_{1}$, the field-level fits to Quijote are significantly larger than the FastPM values and UMF predictions.
    In this comparison, the points plotted for the smaller Quijote box are using a grid with the same physical scale, but a grid resolution that is reduced by a factor of 2 (i.e. for Quijote $N_{g}^{(\mathrm{Quijote})} = 16$). 
  }
    \label{fig:app_quijote}
\end{figure}

As a further test, we measured the Lagrangian and LPNG bias parameters with displacements from \texttt{Gadget} \cite{gadget}.
Specifically, we used the Quijote-PNG \cite{quijote_png}  simulation suite (initial conditions, as well as dark matter particle snapshots and FoF halo catalogs at redshift $z=1$) with $\fnlloc=100$\footnote{Since Quijote-PNG only contains cosmologies with $\fnlloc = \pm 100$ we cannot reproduce the \texttt{FastPM} calculation quantifying the information on $\fnlloc$ performed in Section~\ref{sec:results_fnl}.}.
These simulations were run with a lower particle resolution ($512^{3}$) and in a smaller box ($1~h^{-1}~\mathrm{Gpc}$) than we use in the main text, but employ a tree solver for the small-scale particle-particle interactions, which are not captured by \texttt{FastPM}.
We also use the renormalization procedure described in Section~\ref{app:renorm} for the Quijote-PNG field-level bias measurements, using the publicly avaliable ICs.
This comparison is performed at the same choice of $\Lambda$CDM cosmological parameters, and the Quijote-PNG bias parameters are shown in Figure~\ref{fig:app_quijote}.
There we see that the $\bphi(b_{1})$ relation measured from Quijote-PNG is again qualitatively similar to the SU values (albeit with larger uncertainties), but in detail, the field level values disagree. 
This disagreement between field-level bias values and SU values (as well as the UMF relation) is much larger for Quijote-PNG than for our FastPM measurements.
We speculate that the lower mass halos in both simulations may be affected low mass resolution, and note that doubling the IC particle grid for FastPM leads to changes in inferred $\bphi$ for lower masses (though these changes are smaller than the disagreement with the Quijote points).

\section{Renormalization of quadratic fields}
\label{app:renorm}

We renormalize the quadratic operator fields $\delta_{\Lambda'}^{2}$ and $K_{\Lambda'}^{2}$ to remove the large-scale cutoff-dependent contributions introduced by LPNG \cite{DJS,assassi_renorm_png}
\begin{align}
    [\delta^{2}](\mathbf{k}) &= (\delta_{\Lambda'}^2)(\mathbf{k}) -4 f_{\mathrm{NL}}^{(\mathrm{loc})}\sigma^{2}(\Lambda')\phi(\mathbf{k}),\\
    [K^2](\mathbf{k}) &= (K_{\Lambda'}^2)(\mathbf{k}) -\frac83 f_{\mathrm{NL}}^{(\mathrm{loc})}\sigma^{2}(\Lambda')\phi(\mathbf{k}),
    \label{eqn:quadratic_phi_renorm}
\end{align}
where $\sigma^{2}(\Lambda') = \int_{\mathbf{k}} P_{\delta\delta}(k)W^{2}_{\Lambda'}(\mathbf{k})$.
Failing to account for the renormalization of these terms produces numerically large large-scale contributions to the quadratic operators, which significantly influences the quadratic bias measurements.
We perform the necessary renormalization subtraction in Lagrangian space using a slightly modified version of eqn.~\ref{eqn:quadratic_phi_renorm} on the initial conditions grid before advecting the fields (therefore the $\Lambda'$ in eqn.~\ref{eqn:quadratic_phi_renorm} should be considered distinct from the $\Lambda$ associated with $N_g$ in the rest of this work). 

While the fields involved are all computed on the initial conditions grid, the relation of eqn.~\ref{eqn:quadratic_phi_renorm} was derived under the assumption of a spherically symmetric window $W_{\Lambda'}(k)$ \cite{sjd_quadratic_window_spherical_computation} to perform the angular part of the bispectrum integral (see their Appendix C).
As a result, if we take eqn.~\ref{eqn:quadratic_phi_renorm} at face value, and associate $\sigma^2(\Lambda')$ with the variance of $\delta$ on the IC grid, we do not obtain the value of $\sigma^2$ in  eqn.~\ref{eqn:quadratic_phi_renorm} that removes large-scale cutoff dependence.

One can compute $\sigma(\Lambda')$ without the edge modes beyond $\Lambda'$ that exist on the grid by removing these modes individually.
It would also be possible to use a spherically symmetric filter for this purpose, but a sharp filter in $k-$space introduces artefacts in real space that would impact our field-level bias measurements.
To test whether $\sigma^{2}(\Lambda')$ on the grid was being affected only by the numerics of edge modes, we massively suppressed with a hyperbolic tangent filter with smoothing scale $R=50~\mathrm{Mpc}/h$ and compared the variance on this grid with the theoretical expectation from linear theory for $\sigma^{2}$.
We found that these values agreed to 4 significant figures, indicating that the small-scale modes indeed contributed to a change in $\sigma^{2}(\Lambda')$.

Rather than apply a filter to the grid, we empirically fit two free coefficients $\mathbf{\alpha} = (\alpha_{\delta^2},\alpha_{K^{2}})$ multiplying the component spectra to jointly minimize the large-scale pieces of the non-Gaussian component spectra.
Specifically, the values of $\mathbf{\alpha} = (\alpha_{\delta^2},\alpha_{K^{2}})$ are fitted using Broyden-Fletcher-Goldfarb-Shanno (BFGS) algorithm \cite{NoceWrig06,scipy} $\ell_{2}$ loss normalized by the diagonal Gaussian errors given by the Gaussian component power spectra
\begin{align}
    \label{eqn:min_alpha}
    \alpha_{\delta^2}^{\star} &= \mathrm{argmin}_{\alpha_{\delta^2}} \sum_{k_{i}=k_{f}}^{k_{\mathrm{max}}=0.03~h/\mathrm{Mpc}}  \frac{1}{\sigma^2_{\mathrm{disc.},\delta^2}} \left(\langle \delta^{2}\delta\rangle_{G}(k_i)-\langle \delta^{2}\delta\rangle(k_i) + 4\alpha_{\delta^{2}} \fnlloc \sigma^{2}(\Lambda)\langle \phi \delta\rangle(k_i)\right)^2 \\
    \alpha_{K^2}^{*} &= \mathrm{argmin}_{\alpha_{K^2}} \sum_{k_{i}=k_{f}}^{k_{\mathrm{max}}=0.03~h/\mathrm{Mpc}} \frac{1}{\sigma^2_{\mathrm{disc.},K^2}} \left(\langle K^{2}\delta\rangle_{G}(k_i)-\langle K^{2}\delta\rangle(k_i) + \frac83 \alpha_{K^{2}} \fnlloc \sigma^{2}(\Lambda)\langle \phi \delta\rangle(k_i)\right)^2,
\end{align}
where $\sigma^2_{\mathrm{disc.},\delta^2},\sigma^2_{\mathrm{disc.},K^2}$ are the disconnected covariances for the spectra being fitted.
This process, for example, gives $\alpha_{\delta^2}=0.1$ and $\alpha_{K^{2}}=0.1$ for our bias measurements at $\fnl=100$. 

\section{Priors and profiles}
\label{app:priors}

\begin{figure}[h!]
    \centering
    \includegraphics[width=0.8\textwidth]{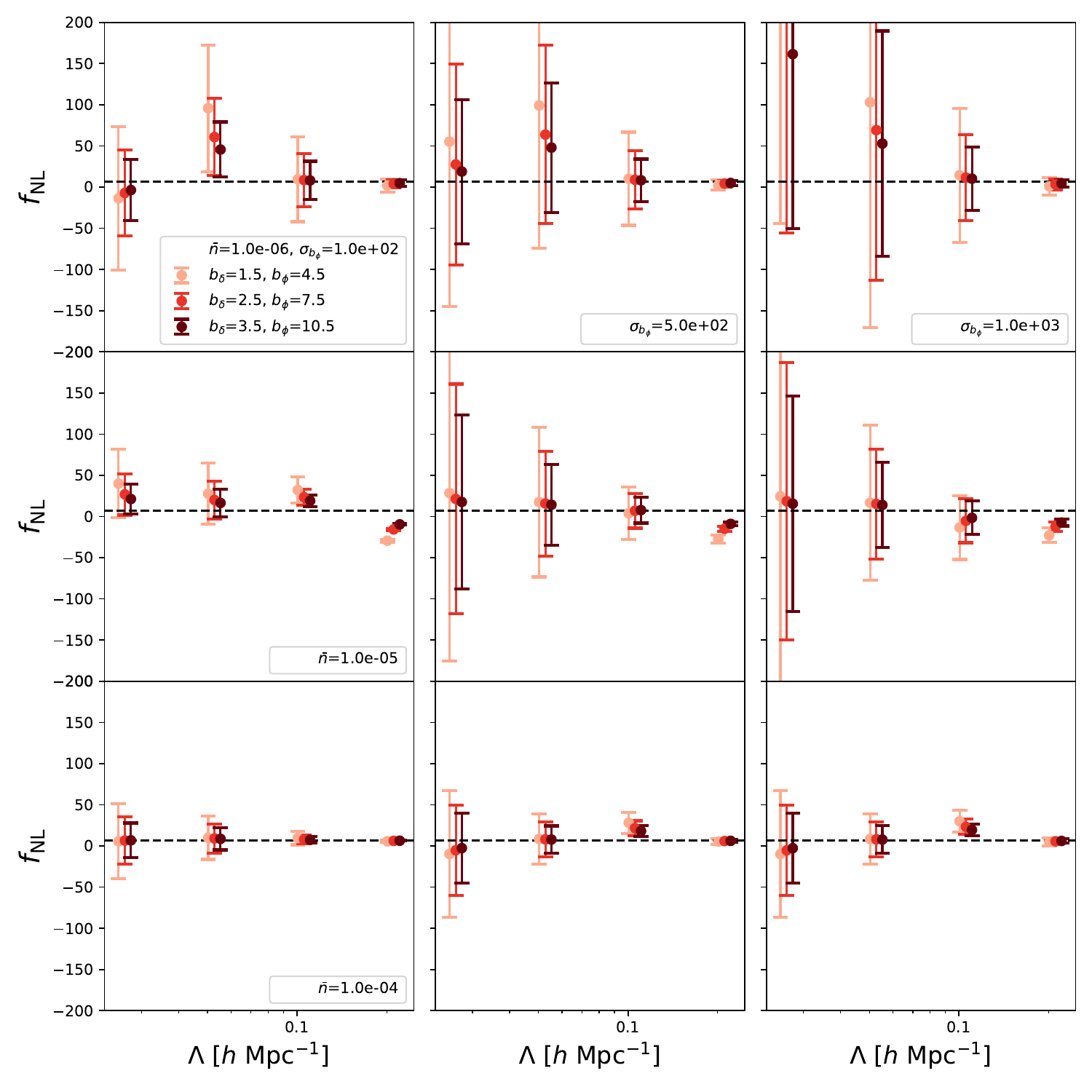} %
    \caption{The impact of both prior strength (columns) and number density (rows) on the profile constraints of $\fnlloc$ (similar to Fig.~\ref{fig:mock_pl_zerotest}, but see the text for differences).
    Different color shades indicate values of $b_\delta$ and $b_{\phi}$ used to generate the perturbative field level mock.
    Rows correspond to different levels of injected Poisson noise corresponding to number densities $\bar{n}$ given in $[h/\mathrm{Mpc}]^3$, while columns correspond to different prior width assumptions for $\bphi$.}
    \label{fig:app_prof_like_tests}
\end{figure}
Figure~\ref{fig:app_prof_like_tests} shows the impact of changing the Poisson noise level (through effective number density in a cell $\bar{n}$) and of adjusting the prior width on $\bphi$.
Here all quadratic bias values are fixed to zero in the mock, but $b_\delta$ is set as in Section~\ref{subsec:fnl_info_mock} and $\bphi$ is set using the UMF relation eqn.~\ref{eqn:umf}.
In this Appendix, we are interested in the impact of priors on statistical constraining power so fix $\fnl$ to the true value used in generating the mock (here $\fnl=7$), but of course a systematic error can bias inferred $\fnl$ results. 
Each color shade corresponds to a different choice of $b_1$ and $b_\phi$ values in the perturbative field-level mock.
Here we consider a linear model, and what is effectively a linear mock.
Moving down a column corresponds to decreasing the injected Poisson noise level, while moving across a row shows the impact of weaker priors.
The prior widths considered here, when small, significantly shrink the size of the $\fnlloc$ constraint uncertainties when the injected Poisson noise is larger (lower $\bar{n}$), though a more modest effect is still visible for lower noise (higher $\bar{n}$)
Clearly, if particularly aggressive yet trusted priors on $\bphi$ are available (e.g. from simulations), a concomitant reduction in $\sigma(\fnlloc)$ is possible.
However, as we discuss in the main text, the linear model used for the illustration in Figure~\ref{fig:app_prof_like_tests} is almost assured to generate statistically biased $\fnlloc$ when applied to realistic tracers.

\bibliographystyle{JHEP}
\bibliography{pngbfl.bib}

\end{document}